%Paper: 9202057
%From: app@cuphyf.phys.columbia.edu (Alexios Polychronakos)
%Date: Mon, 17 Feb 92 15:48:26 EST
%Date (revised): Mon, 17 Feb 92 16:20:35 EST

%%%%%%%%%%THIS IS MACROS AND JNL%%%%%%%%%%%%%%%%%%%%%%%%%%%%%%%%%%%%%%
\def \half {\textstyle {1 \over 2}}

\def \p {{\tilde \pi}}
\def \I {{\tilde I}}

\font\twelverm=cmr10  scaled 1200   \font\twelvei=cmmi10  scaled 1200
\font\twelvesy=cmsy10 scaled 1200   \font\twelveex=cmex10 scaled 1200
\font\twelvebf=cmbx10 scaled 1200   \font\twelvesl=cmsl10 scaled 1200
\font\twelvett=cmtt10 scaled 1200   \font\twelveit=cmti10 scaled 1200
\font\twelvesc=cmcsc10 scaled 1200  \font\twelvesf=cmss10 scaled 1200
\skewchar\twelvei='177   \skewchar\twelvesy='60

%  Define \...point macros to change fonts and spacings consistently

\def\twelvepoint{\normalbaselineskip=12.4pt plus 0.1pt minus 0.1pt
  \abovedisplayskip 12.4pt plus 3pt minus 9pt
  \belowdisplayskip 12.4pt plus 3pt minus 9pt
  \abovedisplayshortskip 0pt plus 3pt
  \belowdisplayshortskip 7.2pt plus 3pt minus 4pt
  \smallskipamount=3.6pt plus1.2pt minus1.2pt
  \medskipamount=7.2pt plus2.4pt minus2.4pt
  \bigskipamount=14.4pt plus4.8pt minus4.8pt
  \def\rm{\fam0\twelverm}          \def\it{\fam\itfam\twelveit}%
  \def\sl{\fam\slfam\twelvesl}     \def\bf{\fam\bffam\twelvebf}%
  \def\mit{\fam 1}                 \def\cal{\fam 2}%
  \def\sc{\twelvesc}               \def\tt{\twelvett}
  \def\sf{\twelvesf}
  \textfont0=\twelverm   \scriptfont0=\tenrm   \scriptscriptfont0=\sevenrm
  \textfont1=\twelvei    \scriptfont1=\teni    \scriptscriptfont1=\seveni
  \textfont2=\twelvesy   \scriptfont2=\tensy   \scriptscriptfont2=\sevensy
  \textfont3=\twelveex   \scriptfont3=\twelveex  \scriptscriptfont3=\twelveex
  \textfont\itfam=\twelveit
  \textfont\slfam=\twelvesl
  \textfont\bffam=\twelvebf \scriptfont\bffam=\tenbf
  \scriptscriptfont\bffam=\sevenbf
  \normalbaselines\rm}

%       tenpoint

%%
%%      Various internal macros
%%

\def\beginlinemode{\endmode
  \begingroup\parskip=0pt \obeylines\def\\{\par}\def\endmode{\par\endgroup}}
\def\beginparmode{\endmode
  \begingroup \def\endmode{\par\endgroup}}
\let\endmode=\par
{\obeylines\gdef\
{}}
\def\singlespace{\baselineskip=\normalbaselineskip}

\def\oneandahalfspace{\baselineskip=\normalbaselineskip
  \multiply\baselineskip by 3 \divide\baselineskip by 2}
\def\doublespace{\baselineskip=\normalbaselineskip \multiply\baselineskip by 2}

\newcount\firstpageno
\firstpageno=2
\footline={\ifnum\pageno<\firstpageno{\hfil}\else{\hfil\twelverm\folio\hfil}\fi}
\def\toppageno{\global\footline={\hfil}\global\headline
  ={\ifnum\pageno<\firstpageno{\hfil}\else{\hfil\twelverm\folio\hfil}\fi}}
\let\rawfootnote=\footnote              % We must set the footnote style
\def\footnote#1#2{{\rm\singlespace\parindent=0pt\parskip=0pt
  \rawfootnote{#1}{#2\hfill\vrule height 0pt depth 6pt width 0pt}}}
\def\raggedcenter{\leftskip=4em plus 12em \rightskip=\leftskip
  \parindent=0pt \parfillskip=0pt \spaceskip=.3333em \xspaceskip=.5em
  \pretolerance=9999 \tolerance=9999
  \hyphenpenalty=9999 \exhyphenpenalty=9999 }
\def\dateline{\rightline{\ifcase\month\or
  January\or February\or March\or April\or May\or June\or
  July\or August\or September\or October\or November\or December\fi
  \space\number\year}}
\def\received{\vskip 3pt plus 0.2fill
 \centerline{\sl (Received\space\ifcase\month\or
  January\or February\or March\or April\or May\or June\or
  July\or August\or September\or October\or November\or December\fi
  \qquad, \number\year)}}

%%
%%      Page layout, margins, font and spacing (feel free to change)
%%

\hsize=6.5truein
\hoffset=0.0truein
\vsize=8.9truein
\voffset=0.0truein
\parskip=\medskipamount
\def\\{\cr}
\twelvepoint            % selects twelvepoint fonts (cf. \tenpoint)
\doublespace            % selects double spacing for main part of paper (cf.
                        %       \singlespace, \oneandahalfspace)
\overfullrule=0pt       % delete the nasty little black boxes for overfull box

%%
%%      This used to be timestamp.tex
%%

\newcount\timehour
\newcount\timeminute
\newcount\timehourminute
\def\daytime{\timehour=\time\divide\timehour by 60
  \timehourminute=\timehour\multiply\timehourminute by-60
  \timeminute=\time\advance\timeminute by \timehourminute
  \number\timehour:\ifnum\timeminute<10{0}\fi\number\timeminute}
\def\today{\number\day\space\ifcase\month\or Jan\or Feb\or Mar
  \or Apr\or May\or Jun\or Jul\or Aug\or Sep\or Oct\or
  Nov\or Dec\fi\space\number\year}

  %  "Draft", Timestamp

%%
%%      The user definitions for major parts of a paper (feel free to change)
%%

      % Preprint number at upper right of title page

\def\cutp#1{
 \rightline{\rm CU--TP--#1}}

       % Preprint number at upper right of title page

\def\title                      %  Title on title page
  {\null\vskip 3pt plus 0.2fill
   \beginlinemode \doublespace \raggedcenter \bf}

\def\author                     %  Author(s) name(s)  on title page
  {\vskip 3pt plus 0.2fill \beginlinemode
   \doublespace \raggedcenter}

\def\affil                      % Affiliations (can intermix with \author)
  {\vskip 3pt plus 0.1fill \beginlinemode
   \oneandahalfspace \raggedcenter \it}

\def\abstract                   % Begin abstract
  {\vskip 3pt plus 0.3fill \beginparmode \narrower
   \oneandahalfspace {\it  Abstract}:\  }

\def\endtopmatter               % End title page, begin body of paper
  {\endpage                     %       This subsumes \body
   \body}

\def\body                       % Begin text body;  can be used to end
  {\beginparmode}               % \title, \author, \affil, \abstract,
                                % \reference, or \figurecaption modes

\def\head#1{                    % Head;  NOTE enclose the text in {}
  \goodbreak\vskip 0.4truein    %  e.g., \head{I. Introduction}
  {\immediate\write16{#1}
   \raggedcenter {\sc #1} \par }
   \nobreak\vskip 0truein\nobreak}

\def\beneathrel#1\under#2{\mathrel{\mathop{#2}\limits_{#1}}}

\def\refto#1{$^{#1}$}           % For references in text as superscript

\def\references                 % Begin references -- basic format is Phys Rev
  {\head{References}            % i.e., volume, page, year (space after
% commas).
   \beginparmode
   \frenchspacing \parindent=0pt    %\leftskip=1truecm ?
   \parskip=0pt \everypar{\hangindent=20pt\hangafter=1}}

\gdef\refis#1{\item{#1.\ }}                     % Ref list numbers.

\gdef\journal#1,#2,#3,#4.{              % Journal reference.  Comma sets
    {\sl #1~}{\bf #2}, #3 (#4).}                % off: name, vol, page, year

\def\endreferences{\body}

\def\figurecaptions             % Begin figure captions
  {\endpage
   \beginparmode
   \head{Figure Captions}
}

\def\endpage                    %  Eject a page
  {\vfill\eject}

\def\endpaper                   %  Ways to say goodbye
  {\endmode\vfill\supereject}

%%
%%      Various little user definitions
%%

\def\ref#1{Ref.~#1}                     %       for inline references
\def\Ref#1{Ref.~#1}                     %       ditto
\def\[#1]{[\cite{#1}]}
\def\cite#1{{#1}}
          % For citation of equation numbers
        %       ditto
                     %       ditto
                   %       ditto
\def\(#1){(\call{#1})}
\def\call#1{{#1}}
\def\taghead#1{}
\def\frac#1#2{{#1 \over #2}}
\def\half{{\frac 12}}

\def\12{{1\over2}}

\def\sla{\raise.15ex\hbox{$/$}\kern-.57em}
\def\leaderfill{\leaders\hbox to 1em{\hss.\hss}\hfill}
\def\twiddle{\lower.9ex\rlap{$\kern-.1em\scriptstyle\sim$}}
\def\bigtwiddle{\lower1.ex\rlap{$\sim$}}
\def\gtwid{\mathrel{\raise.3ex\hbox{$>$\kern-.75em\lower1ex\hbox{$\sim$}}}}
\def\ltwid{\mathrel{\raise.3ex\hbox{$<$\kern-.75em\lower1ex\hbox{$\sim$}}}}
\def\square{\kern1pt\vbox{\hrule height 1.2pt\hbox{\vrule width 1.2pt\hskip 3pt
   \vbox{\vskip 6pt}\hskip 3pt\vrule width 0.6pt}\hrule height 0.6pt}\kern1pt}
\def\tdot#1{\mathord{\mathop{#1}\limits^{\kern2pt\ldots}}}

\def\pmb#1{\setbox0=\hbox{#1}%
  \kern-.025em\copy0\kern-\wd0
  \kern  .05em\copy0\kern-\wd0
  \kern-.025em\raise.0433em\box0 }

\catcode`@=11
\newcount\tagnumber\tagnumber=0

\immediate\newwrite\eqnfile
\newif\if@qnfile\@qnfilefalse
\def\write@qn#1{}
\def\writenew@qn#1{}
\def\w@rnwrite#1{\write@qn{#1}\message{#1}}
\def\@rrwrite#1{\write@qn{#1}\errmessage{#1}}

\def\taghead#1{\gdef\t@ghead{#1}\global\tagnumber=0}
\def\t@ghead{}

\expandafter\def\csname @qnnum-3\endcsname
  {{\t@ghead\advance\tagnumber by -3\relax\number\tagnumber}}
\expandafter\def\csname @qnnum-2\endcsname
  {{\t@ghead\advance\tagnumber by -2\relax\number\tagnumber}}
\expandafter\def\csname @qnnum-1\endcsname
  {{\t@ghead\advance\tagnumber by -1\relax\number\tagnumber}}
\expandafter\def\csname @qnnum0\endcsname
  {\t@ghead\number\tagnumber}
\expandafter\def\csname @qnnum+1\endcsname
  {{\t@ghead\advance\tagnumber by 1\relax\number\tagnumber}}
\expandafter\def\csname @qnnum+2\endcsname
  {{\t@ghead\advance\tagnumber by 2\relax\number\tagnumber}}
\expandafter\def\csname @qnnum+3\endcsname
  {{\t@ghead\advance\tagnumber by 3\relax\number\tagnumber}}

\def\equationfile{%
  \@qnfiletrue\immediate\openout\eqnfile=\jobname.eqn%
  \def\write@qn##1{\if@qnfile\immediate\write\eqnfile{##1}\fi}
  \def\writenew@qn##1{\if@qnfile\immediate\write\eqnfile
    {\noexpand\tag{##1} = (\t@ghead\number\tagnumber)}\fi}
}

\def\callall#1{\xdef#1##1{#1{\noexpand\call{##1}}}}
\def\call#1{\each@rg\callr@nge{#1}}

\def\each@rg#1#2{{\let\thecsname=#1\expandafter\first@rg#2,\end,}}
\def\first@rg#1,{\thecsname{#1}\apply@rg}
\def\apply@rg#1,{\ifx\end#1\let\next=\relax%
\else,\thecsname{#1}\let\next=\apply@rg\fi\next}

\def\callr@nge#1{\calldor@nge#1-\end-}
\def\callr@ngeat#1\end-{#1}
\def\calldor@nge#1-#2-{\ifx\end#2\@qneatspace#1 %
  \else\calll@@p{#1}{#2}\callr@ngeat\fi}
\def\calll@@p#1#2{\ifnum#1>#2{\@rrwrite{Equation range #1-#2\space is bad.}
\errhelp{If you call a series of equations by the notation M-N, then M and
N must be integers, and N must be greater than or equal to M.}}\else%
 {\count0=#1\count1=#2\advance\count1
by1\relax\expandafter\@qncall\the\count0,%
  \loop\advance\count0 by1\relax%
    \ifnum\count0<\count1,\expandafter\@qncall\the\count0,%
  \repeat}\fi}

\def\@qneatspace#1#2 {\@qncall#1#2,}
\def\@qncall#1,{\ifunc@lled{#1}{\def\next{#1}\ifx\next\empty\else
  \w@rnwrite{Equation number \noexpand\(>>#1<<) has not been defined yet.}
  >>#1<<\fi}\else\csname @qnnum#1\endcsname\fi}

\let\eqnono=\eqno
\def\eqno(#1){\tag#1}
\def\tag#1$${\eqnono(\displayt@g#1 )$$}

\def\aligntag#1\endaligntag
  $${\gdef\tag##1\\{&(##1 )\cr}\eqalignno{#1\\}$$
  \gdef\tag##1$${\eqnono(\displayt@g##1 )$$}}

\def\eqalignno#1{\displ@y \tabskip\centering
  \halign to\displaywidth{\hfil$\displaystyle{##}$\tabskip\z@skip
    &$\displaystyle{{}##}$\hfil\tabskip\centering
    &\llap{$\displayt@gpar##$}\tabskip\z@skip\crcr
    #1\crcr}}

\def\displayt@gpar(#1){(\displayt@g#1 )}

\def\displayt@g#1 {\rm\ifunc@lled{#1}\global\advance\tagnumber by1
        {\def\next{#1}\ifx\next\empty\else\expandafter
        \xdef\csname @qnnum#1\endcsname{\t@ghead\number\tagnumber}\fi}%
  \writenew@qn{#1}\t@ghead\number\tagnumber\else
        {\edef\next{\t@ghead\number\tagnumber}%
        \expandafter\ifx\csname @qnnum#1\endcsname\next\else
        \w@rnwrite{Equation \noexpand\tag{#1} is a duplicate number.}\fi}%
  \csname @qnnum#1\endcsname\fi}

\def\ifunc@lled#1{\expandafter\ifx\csname @qnnum#1\endcsname\relax}

\let\@qnend=\end\gdef\end{\if@qnfile
\immediate\write16{Equation numbers written on []\jobname.EQN.}\fi\@qnend}

\catcode`@=12

\catcode`@=11
\newcount\r@fcount \r@fcount=0
\newcount\r@fcurr
\immediate\newwrite\reffile
\newif\ifr@ffile\r@ffilefalse
\def\w@rnwrite#1{\ifr@ffile\immediate\write\reffile{#1}\fi\message{#1}}

\def\writer@f#1>>{}
\def\referencefile{%                      Stuff to write .REF file
  \r@ffiletrue\immediate\openout\reffile=\jobname.ref%
  \def\writer@f##1>>{\ifr@ffile\immediate\write\reffile%
    {\noexpand\refis{##1} = \csname r@fnum##1\endcsname = %
     \expandafter\expandafter\expandafter\strip@t\expandafter%
     \meaning\csname r@ftext\csname r@fnum##1\endcsname\endcsname}\fi}%
  \def\strip@t##1>>{}}

\def\citeall#1{\xdef#1##1{#1{\noexpand\cite{##1}}}}
\def\cite#1{\each@rg\citer@nge{#1}}     % Variable No. of args, separated by
% ","

\def\each@rg#1#2{{\let\thecsname=#1\expandafter\first@rg#2,\end,}}
\def\first@rg#1,{\thecsname{#1}\apply@rg}       % each@ag is a general purpose
\def\apply@rg#1,{\ifx\end#1\let\next=\relax%      variable no. of arg. macro.
\else,\thecsname{#1}\let\next=\apply@rg\fi\next}% args separated by commas

\def\citer@nge#1{\citedor@nge#1-\end-}  % Check for M-N range (M and N numbers)
\def\citer@ngeat#1\end-{#1}
\def\citedor@nge#1-#2-{\ifx\end#2\r@featspace#1 % Single argument
  \else\citel@@p{#1}{#2}\citer@ngeat\fi}        % M-N range of arguments
\def\citel@@p#1#2{\ifnum#1>#2{\errmessage{Reference range #1-#2\space is bad.}
    \errhelp{If you cite a series of references by the notation M-N, then M and
    N must be integers, and N must be greater than or equal to M.}}\else%
 {\count0=#1\count1=#2\advance\count1
by1\relax\expandafter\r@fcite\the\count0,%
  \loop\advance\count0 by1\relax%         Loop from M to N
    \ifnum\count0<\count1,\expandafter\r@fcite\the\count0,%
  \repeat}\fi}

\def\r@featspace#1#2 {\r@fcite#1#2,}    % Eat spaces at beginning or end of arg
\def\r@fcite#1,{\ifuncit@d{#1}          % Cite individual reference
    \expandafter\gdef\csname r@ftext\number\r@fcount\endcsname%
    {\message{Reference #1 to be supplied.}\writer@f#1>>#1 to be supplied.\par
     }\fi%
  \csname r@fnum#1\endcsname}

\def\ifuncit@d#1{\expandafter\ifx\csname r@fnum#1\endcsname\relax%
\global\advance\r@fcount by1%
\expandafter\xdef\csname r@fnum#1\endcsname{\number\r@fcount}}

\let\r@fis=\refis                       % Save old \refis, redefine
\def\refis#1#2#3\par{\ifuncit@d{#1}%      Use two params #2 #3 to strip blank
    \w@rnwrite{Reference #1=\number\r@fcount\space is not cited up to now.}\fi%
  \expandafter\gdef\csname r@ftext\csname r@fnum#1\endcsname\endcsname%
  {\writer@f#1>>#2#3\par}}

\def\r@ferr{\endreferences\errmessage{I was expecting to see
\noexpand\endreferences before now;  I have inserted it here.}}
\let\r@ferences=\references
\def\references{\r@ferences\def\endmode{\r@ferr\par\endgroup}}

\let\endr@ferences=\endreferences
\def\endreferences{\r@fcurr=0%            Save old \endreferences, redefine
  {\loop\ifnum\r@fcurr<\r@fcount%         Loop over refnum and produce text
    \advance\r@fcurr by 1\relax\expandafter\r@fis\expandafter{\number\r@fcurr}%
    \csname r@ftext\number\r@fcurr\endcsname%
  \repeat}\gdef\r@ferr{}\endr@ferences}

% Save old \endpaper, redefine it to write parting message.

\let\r@fend=\endpaper\gdef\endpaper{\ifr@ffile
\immediate\write16{Cross References written on []\jobname.REF.}\fi\r@fend}

\catcode`@=12

\citeall\refto          % These macros will generate citations
\citeall\ref            %
\citeall\Ref            %
%%
%%              TABLEOFC.TEX                    6/2/85  Doug Eardley
%%
%%              Used with JNL.TEX to produce a table of contents. To use,
%%      \input this file after JNL.  After running TEX <filename>,
%%      a new TeX file <filename>.TOC will be written, which contains the
%%      table of contents.  The latter file can then be edited (if necessary)
%%      and TEX'd itself.  All items #1 which appear in \head{#1} are listed.
%%              To list other items, use the macro \tocitem{#1}.
%%      To list automatically all items which appear in some macro \macroname,
%%      declare \tocitemall\macroname.  Ditto \tocitemitem and \tocitemitemall
%%      for subitems, and \tocitemitemitem and \tocitemitemitemall for subsub-
%%      items.  The macro \macroname must have one exactly one argument #1.
%%              At present, TABLEOFC is ***incompatible*** with PPT.

\catcode`@=11
\newwrite\tocfile\openout\tocfile=\jobname.toc
\newlinechar=`^^J
\write\tocfile{\string\input\space jnl^^J
  \string\pageno=-1\string\firstpageno=-1000\string\singlespace
  \string\null\string\vfill\string\centerline{TABLE OF CONTENTS}^^J
  \string\vskip 0.5 truein\string\rightline{\string\underbar{Page}}\smallskip}

\def\tocitem#1{%                Add item to table of contents, this page
  \t@cskip{#1}\bigskip}
\def\tocitemitem#1{%            Ditto subitem
  \t@cskip{\quad#1}\medskip}
\def\tocitemitemitem#1{%        Ditto subsubitem
  \t@cskip{\qquad#1}\smallskip}
\def\tocitemall#1{%             Make all macro#1 tocitem's
  \xdef#1##1{#1{##1}\noexpand\tocitem{##1}}}
\def\tocitemitemall#1{%         Make all macro#1 tocitemitem's
  \xdef#1##1{#1{##1}\noexpand\tocitemitem{##1}}}
\def\tocitemitemitemall#1{%             Make all macro#1 tocitemitemitem's
  \xdef#1##1{#1{##1}\noexpand\tocitemitemitem{##1}}}

\def\t@cskip#1#2{%              Add item to T.O.C with skip, this page
  \write\tocfile{\string#2\string\line{^^J
  #1\string\leaderfill\space\number\folio}}}

%\def\tocitempage#1#2{%         Add item to table of contents on page #2
%  \write\tocfile{\string\bigskip\string\line{^^J
%  #1\string\leaderfill\space\number#2}}}
%
%\newcount\pastpageno\pastpageno=0
%\def\tocitempast#1#2{%         Add item to toc, #2 pages past end of paper;
%  \ifnum\pastpageno=0
%    \global\pastpageno=\pageno
%    \global\advance\pastpageno by 1 \fi
%  \tocitempage{#1}\pastpageno \global\advance\pastpageno by #2}

\def\t@cproduce{%                 Spit out T.O.C.
  \write\tocfile{\string\vfill\string\vfill\string\supereject\string\end}
  \closeout\tocfile
  \immediate\write16{Table of Contents written on []\jobname.TOC.}}

% Save old \endpaper, redefine it to write .TOCfile.

\let\t@cend=\endpaper\def\endpaper{\t@cproduce\t@cend}

\catcode`@=12

\tocitemall\head                % \head{} will now generate T.O.C. entries

%%%%%%%%%%%%%%PAPER STARTS HERE%%%%%%%%%%%%%%%%%%%%%%%%%%%%%%%%%%%%%

\cutp{551}
\title{Exchange Operator Formalism for Integrable Systems of Particles}
\author{\bf Alexios P. Polychronakos}
\affil{Pupin Physics Laboratories, Columbia University,
New York, NY 10027}
\abstract{We formulate one dimensional many body integrable systems in
terms of a new set of phase space variables involving exchange operators.
The hamiltonian in these variables assumes a decoupled form.
This greatly simplifies the derivation of the conserved charges and
the proof of their commutativity at the quantum level.}

\endtopmatter

\body
\baselineskip=20pt

In one spatial dimension a class of integrable many-body systems is
known, referred to as the Calogero-Sutherland-Moser systems$^{1-3}$. They
constitute of many identical nonrelativistic particles interacting
through two-body potentials of the inverse square type and its
generalizations, namely the inverse sine square and the Weierstrass
two-body potentials. These models are related to root systems of
$A_n$ algebras$^4$. Corresponding systems related to root systems of
other algebras exist, but their two-body potentials are not translationally
and/or permutation invariant$^5$. We will restrict ourselves to the
$A_n$ systems. For a comprehensive review of these systems see ref$.$ 5.

Many of the above systems admit a matrix formulation$^{5,6}$. Using this
formulation, a generalization of these systems was found recently
where the particles also feel external potentials of particular types$^7$.
These systems, apart from their purely mathematical interest, are
also of significant physical interest, since they are relevant to
fractional statistics and anyons$^8$, spin chain models$^9$, soliton wave
propagation$^{10}$ and, indirectly, to nonperturbative two-dimensional
quantum gravity$^{11}$.

The purpose of this paper is to present an ``exchange operator"
formalism for these systems which renders their integrable structure
explicit. Specifically, we will write generalized momentum operators
in terms of which the integrals of motion assume a ``decoupled" form.
This will allow for an easy proof of commutativity at the quantum level.

Let $\{ x_i , p_i \}$, $i=1, \dots N$ be the coordinates and momenta
of $N$ one-dimensional quantum mechanical particles, obeying canonical
commutation relations, and let $M_{ij}$ be the particle permutation
operators, obeying
$$
M_{ij} = M_{ji} = M_{ij}^\dagger \,\,,\,\,\,\, M_{ij}^2 = 1
\eqno(a1)$$
$$
M_{ij} A_j = A_i M_{ij} \,\,,\,\,\,\,
M_{ij} A_k = A_k M_{ij} \,\,,\,\,\,\, {\rm for} ~ k \neq i,j
\eqno(a)$$
where $A_i$ is any operator (including $M_{ij}$ themselves) carrying
one or more particle indices. Then define the ``coupled" momentum operators
$$
\pi_i = p_i + i \sum_{j \neq i} V_{ij} M_{ij} \,\,,\,\,\,\,
V_{ij} \equiv V( x_i - x_j )
\eqno(b)$$
with $V(x)$ an as yet undetermined function. Note that the $\pi_i$ are
``good" one-particle operators, that is they satisfy \(a), since the
remaining particle indices in \(b) appear in a permutation symmetric way.
If we impose the hermiticity condition on $\pi_i$
$$
\pi_i = \pi_i^\dagger
\eqno(c)$$
then $V(x)$ must obey
$$
V(x)^\dagger = - V(-x)
\eqno(d)$$
Consider now a hamiltonian for the system which takes a free form
in terms of $\pi_i$'s, that is,
$$
H = \half \sum_i \pi_i^2
\eqno(e)$$
In terms of the original phase space variables, $H$ takes the form
$$
H = \half \sum_i p_i^2 +
\half \sum_{i \neq j} \left [ i V_{ij} ( p_i + p_j ) M_{ij}
+ V^\prime_{ij} M_{ij} + V_{ij}^2 \right ]
- {1 \over 6} \sum_{i \neq j \neq k \neq i} V_{ijk} M_{ijk}
\eqno(f)$$
In the above, $V^\prime (x)$ is the derivative of $V(x)$ and we
defined
$$
V_{ijk} = V_{ij} V_{jk} + V_{jk} V_{ki} + V_{ki} V_{ij}
\eqno(g)$$
$M_{ijk}$ is the generator of cyclic permutations in three indices,
that is,
$$
M_{ijk} = M_{jki} = M_{kij} = M_{jik}^\dagger = M_{ij} M_{jk}
\eqno(h)$$
If we demand that the above expression for $H$ become the sum of an
ordinary kinetic term and potential terms, the terms linear in $p_i$
should drop, and this will happen if
$$
V(-x) = -V(x)
\eqno(k)$$
Finally, if we want the above hamiltonian to contain only two-body
potentials, the function $V(x)$ should satisfy
$$
V(x) V(y) + V(y) V (z) + V(z) V(x) = W(x) + W(y) + W(z)
\,\,,\,\,\,\, {\rm for} ~ x+y+z = 0
\eqno(l)$$
where $W(x)$ is a new symmetric function. $H$ takes then the form
$$
H = \half \sum_i p_i^2 +
\sum_{i < j} \left [ V_{ij}^2  + V^\prime_{ij} M_{ij}
- W_{ij} \sum_{k \neq i,j} M_{ijk} \right ]
\eqno(m)$$
and the commutator of $\pi$'s is evaluated to be
$$
[ \pi_i , \pi_j ] = \sum_{k \neq i,j} V_{ijk} [ M_{ijk} - M_{jik} ]
\eqno(n)$$

Eq$.$ \(l) is a well-known
functional equation for $V$ which also emerges as a condition for
factorizability of the ground state of many-body systems$^3$. It can be
readily solved through a small-$x$ expansion and all its solutions are
available$^{12}$. Here we consider in sequence the solutions of most interest.

Assume first that $W(x) = 0$. Then \(l) is solved by
$$
V(x) = {l \over x}
\eqno(p)$$
with $l$ a real parameter, and the hamiltonian takes the form
$$
H = \half \sum_i p_i^2 + \sum_{i > j} {l (l - M_{ij} ) \over ( x_i
- x_j )^2 }
\eqno(q)$$
Define the totally symmetric quantities
$$
I_n = \sum_i \pi_i^n
\eqno(o)$$
Since in this case $V_{ijk} = 0$ we see from \(n) that the $\pi_i$
{\it commute} and
therefore the $I_n$ also commute. Moreover, since they commute with
all $M_{ij}$, their projections in the bosonic or fermionic subspaces of
the Hilbert space also commute. In these subspaces the $M_{ij}$ simply
become $\pm 1$ and $\half I_2 = H$ becomes the hamiltonian of a set
of particles interacting through inverse square potentials of strength
$l(l \mp 1)$. Further, the higher quantities $I_n$ projected in these
subspaces become the integrals of motion of the above hamiltonian.
Therefore these integrals commute in these subspaces.

To show that these integrals commute in the full Hilbert space it
suffices to notice that they are local operators, since they involve
derivatives of at most $n$-th degree. To know their action on the
wavefunction at any point it suffices to know the wavefunction in a small
neighborhood around that point. Therefore, the fact that they commute
cannot depend on global information on the wavefunction, namely its
symmetry or antisymmetry. Thus, if they commute for bosonic or fermionic
states they must commute unconditionally.

In the case $W(x) = \,$constant, the solution for $V$ is
$$
V(x) = {l \cot ax} \,\,,\,\,\,\, {\rm or}\,\,\,\, V(x) = {l \coth ax}
\eqno(pp)$$
depending on the sign of the constant. Choosing the positive sign,
and making $a=1$ by appropriate choice of units, we have in subspaces
of definite symmetry
$$
H = \half \sum_i p_i^2 + \sum_{i<j} {l(l \mp 1) \over \sin^2 ( x_i
- x_j )} - l^2 {N(N^2 -1) \over 6}
\eqno(qq)$$
This is the Sutherland model of particles interacting through inverse
sine square potentials. In this case $V_{ijk} = l^2$ and thus the
$\pi_i$'s do not commute. To show the existence of conserved quantities,
define the new operators
$$
\p_i = \pi_i + l \sum_{j \neq i} M_{ij}
\eqno(r)$$
which are also ``good" one-particle operators and obey the
commutation relations
$$
[ \p_i , \p_j ] = 2l ( \p_i M_{ij} - M_{ij} \p_i )
\eqno(s)$$
The corresponding conserved quantities $\I_n$ constructed
from $\p_i$ can be shown to commute as follows:
$$
\eqalign{
[ \p_i^n , \p_j ] &= \sum_{\alpha = 0}^{n-1} \, \p_i^\alpha
[ \p_i , \p_j ] \, \p_i^{n-\alpha -1} \cr
&= 2l ( \p_i^n M_{ij} - M_{ij} \p_i^n ) = 2l ( M_{ij} \p_j^n -
\p_j^n M_{ij} ) \cr }
\eqno(t)$$
and thus
$$
\eqalign{
[ \I_n , \I_m ] &= \sum_{i,j} [ \p_i^n , \p_j^m ]
= \sum_{i,j} \sum_{\alpha = 0}^{m-1} \p_j^\alpha \, [ \p_i^n ,
\p_j ] \p_j^{m-\alpha -1} \cr
&= 2l \sum_{i,j} \sum_{\alpha = 0}^{m-1} \left( \p_j^\alpha M_{ij}
\p_j^{m+n-\alpha -1} - \p_j^{\alpha +n} M_{ij} \p_j^{m-\alpha- 1}
\right) \cr
&= 2l \sum_{i,j} \left( \sum_{\alpha = 0}^{m-1} - \sum_{\alpha = n}^{m+n-1}
\right) \p_j^\alpha M_{ij} \p_j^{m+n-\alpha -1} \cr }
\eqno(u)$$
Antisymmetrizing \(u) explicitly in $n$ and $m$, we get
$$
[ \I_n , \I_m ] = l \sum_{i,j} \left( \sum_{\alpha = 0}^{m-1} -
\sum_{\alpha = n}^{m+n-1} - \sum_{\alpha = 0}^{n-1} +
\sum_{\alpha = m}^{m+n-1} \right) \p_j^\alpha M_{ij}
\p_j^{m+n-\alpha -1} = 0
\eqno(v)$$
Therefore the $\I_n$ commute. In subspaces of definite symmetry,
on the other hand, they reduce to combinations of $I_n$, e.g.,
$$
\I_1 = I_1 \mp l N(N-1) \,\,,\,\,\,\,
\I_2 = I_2 \mp 2l (N-1) I_1 + l^2 N(N-1)^2 \,\,,\,\,\,\,{\rm e.t.c.}
\eqno(w)$$
Therefore the $I_n$ commute as well. By repeating the above argument,
or simply by analytic continuation, we can also deal with the $\coth ax$
solution in \(pp) which corresponds to the inverse hyperbolic sine
square potential.

A singular solution of \(l) for $W(x)$ a negative constant is
$$
V(x) = l \, {\rm sign} (x)
\eqno(ww)$$
which leads to the well-known system of particles with mutual
delta-function potentials$^{13}$. It can be treated as above, with
some extra care to possible singularities.

Finally, consider the operators
$$
h_i = ( \pi_i + i \omega x_i ) ( \pi_i - i \omega x_i )
\equiv a_i^\dagger a_i
\eqno(x)$$
for $V(x)$ as in \(p). Using the
commutativity of the $\pi_i$ in this case as well as
$$
[ x_i , \pi_j ] = i \delta_{ij} \bigl( 1 + l \sum_{k \neq i} M_{ik}
\bigr) - i (1- \delta_{ij} ) l M_{ij}
\eqno(yy)$$
we find
$$
[ a_i , a_j ] = [ a_i^\dagger , a_j^\dagger ] = 0
$$
$$
[ a_i , a_j^\dagger ] = -2l \omega M_{ij} ~~~ {\rm (for ~ i \neq j)}
\eqno(y)$$
and thus
$$
[ h_i , h_j ] = -2l \omega ( h_i M_{ij} - M_{ij} h_i )
\eqno(z)$$
We observe that the commutation relations of the $h_i$ are similar to the
ones of $\p_i$ in \(s). Therefore the quantities $I_n$ defined now
$$
I_n = \sum_i h_i^n
\eqno(aa)$$
can be shown to commute in a way similar to the one for the $\I_n$.
In particular, the hamiltonian $H = \half I_1$ in the bosonic or fermionic
subspace becomes
$$
H = \half \sum_i p_i^2 + \sum_{i > j} {l (l \mp 1 )
\over ( x_i - x_j )^2 } + \half \sum_i \omega^2 x_i^2
-N {\omega \over 2} \pm l {N(N-1) \over 2} \omega
\eqno(bb)$$
This is the Calogero model of harmonic plus inverse square potentials,
and we have derived its integrals of motion.

Notice that the constant terms appearing in \(qq) and \(bb) are the
negative of the ground state energy of the corresponding hamiltonians,
thus shifting the ground state energy to zero. In fact we can easily
find the ground state wavefunction noticing that the above $H$'s are
positive definite and thus if we can find states $\psi_{_S}$ and
$\psi_{_C}$ satisfying
$$
\pi_i \psi_{_S} = 0 \,\,,\,\,\,\, {\rm or} \,\,\,\,
a_i \psi_{_C} = 0
\eqno(cc)$$
these will be the ground state. Taking them further to be bosonic,
they thus must satisfy
$$
{\partial_i \psi_{_S} \over \psi_{_S}} = \sum_{j \neq i} l \cot ( x_i - x_j )
\eqno(dd)$$
or
$$
{\partial_i \psi_{_C} \over \psi_{_C}} = - \omega x_i + \sum_{j \neq i}
{l \over x_i - x_j }
\eqno(ee)$$
respectively. By integrating \(dd) and \(ee) we easily find the
Sutherland and Calogero ground state wavefunctions
$$
\psi_{_S} = \prod_{i<j} | \sin ( x_i - x_j ) |^l \,\,,\,\,\,\,
\psi_{_C} = \prod_{i<j} | x_i - x_j |^l \, e^{- \half \omega \sum_i x_i^2}
\eqno(ff)$$

In conclusion, we see that the above formalism identifies a better
set of phase space ``momentum" variables, which allow for an effortless
and relatively straightforward derivation of the integrability of
these systems. It is also remarkable that the above proofs work
directly in the quantum regime (the exchange operators $M_{ij}$
have no classical counterpart), thus circumventing the operator
ordering problems encountered when constructing the quantum
integrals of motion starting from the classical standpoint. It is
hoped that this formalism will provide an easy proof of the quantum
integrability of the systems recently found in ref$.$ 7, or
even that it will lead to as yet uncovered new integrable systems.
We hope to come back to these issues in a future publication.

This research was supported in part by a United States Department
of Energy grant.

\bigskip
\bigskip
\centerline {\bf REFERENCES}

\noindent
\item{$^1$}
F. Calogero, {\it J. Math. Phys.} {\bf 10}, 2191 and 2197 (1969) and
{\bf 12}, 419 (1971).

\item{$^2$}
J. Moser, {\it Adv. Math.} {\bf 16}, 1 (1975); F. Calogero, {\it Lett.
Nuovo Cim.} {\bf 13}, 411 (1975); F. Calogero and C. Marchioro, {\it Lett.
Nuovo Cim.} {\bf 13}, 383 (1975).

\item{$^3$}
B. Sutherland, {\it Phys. Rev.} {\bf A4}, 2019 (1971) and {\bf A5},
1372 (1972); {\it Phys. Rev. Lett.} {\bf 34}, 1083 (1975).

\item{$^4$}
M.A. Olshanetsky and A.M. Perelomov, {\it Invent. Math.} {\bf 37},
93 (1976).

\item{$^5$}
M.A. Olshanetsky and A.M. Perelomov, {\it Phys. Rep.} {\bf 71}, 314 (1981)
and {\bf 94}, 6 (1983).

\item{$^6$}
D. Kazdan, B. Kostant and S. Sternberg, {\it Comm. Pure Appl.
Math.} {\bf 31}, 481 (1978).

\item{$^7$}
A.P. Polychronakos, {\it Phys. Lett.} {\bf B266}, 29 (1991); Columbia
preprints CU-TP-527, July 1991, to appear in Phys. Lett. B and
CU-TP-537, October 1991, to appear in Phys. Lett. B.

\item{$^8$}
J.M. Leinaas and J. Myrheim, {\it Phys. Rev.} {\bf B37}, 9286 (1988);
A.P. Polychronakos, {\it Nucl. Phys.} {\bf B324}, 597 (1989) and
{\it Phys. Lett.} {\bf B264}, 362 (1991).

\item{$^9$}
F.D.M. Haldane, {\it Phys. Rev. Lett.} {\bf 60}, 635 (1988) and
{\bf 66}, 1529 (1991); B.S. Shastry, {\it Phys. Rev. Lett.}
{\bf 60}, 639 (1988); F. Gebhard and A.E. Ruckenstein,
{\it Phys. Rev. Lett.} {\bf 68}, 244 (1992).

\item{$^{10}$}
H.H. Chen, Y.C. Lee and N.R. Pereira, {\it Phys. Fluids}
{\bf 22}(1) , 187 (1979).

\item{$^{11}$}
V.A. Kazakov, {\it Random Surfaces and Quantum Gravity},
Cargese Lectures 1990, eds. O. Alvarez {\it et. al.}

\item{$^{12}$}
F. Calogero, {\it Lett. Nuovo Cim.} {\bf 13}, 507 (1975).

\item{$^{13}$}
E.H. Lieb and W. Liniger, {\it Phys. Rev.} {\bf 130}, 1605 (1963);
C.N. Yang, {\it Phys. Rev. Lett.} {\bf 19}, 1312 (1967) and
{\it Phys. Rev.} {\bf 168}, 1920 (1968).

\end